\documentclass[10pt,twocolumn,letterpaper]{article}

\usepackage{times}
\usepackage{epsfig}
\usepackage{graphicx}
\usepackage{amsmath}
\usepackage{amssymb}

\usepackage[breaklinks=true,bookmarks=false]{hyperref}

\def\cvprPaperID{****} 

\begin{document}

\title{Fault Detection in Ball Bearings}

\author{Joshua Pickard\\
{\tt\small jpic@umich.edu}
\and
Sarah Moll\\
{\tt\small srmoll@umich.edu}
}
\date{May 2021}
\maketitle

\begin{abstract}
Ball bearing joints are a critical component in all rotating machinery, and detecting and locating faults in these joints is a significant problem in industry and research. Intelligent fault detection (IFD) is the process of applying machine learning and other statistical methods to monitor the health states of machines. This paper explores the construction of vibration images, a preprocessing technique that has been previously used to train convolutional neural networks for ball bearing joint IFD. The main results demonstrate the robustness of this technique by applying it to a larger dataset than previously used and exploring the hyperparameters used in constructing the vibration images.
\end{abstract}

\section{Introduction}

Intelligent fault diagnosis (IFD) is the application of machine learning methodologies to monitor the health states of machines. The focus of this paper is the use of a unique preprocessing method in conjunction with convolutional neural networks for the purpose of IFD on ball bearing joints, a critical component in rotating machinery. 

Detecting and locating faults in bearing joints is an important problem in industry and research. From cars to windmills, bearing joints are a critical component in machines used in everyday life, and it is estimated that faults in these joints, either from poor manufacturing practices or overuse, accounts for 40-50\% of all motor failures [1]. Not only does this limit the utility of a machine, but it can put operators of these machines in danger. Automating the detection of faults in ball bearings has the potential to reduce downtime and improve safety in multiple industries.

From a research perspective, ball bearing IFD draws on knowledge from many fields ranging from signals processing to deep learning. The most common input for a model for bearing IFD is a vibration signal collected from near the bearing joint. This paper and the related work use these signals to construct 2D representations called vibration images, which are classified using standard machine learning techniques.

\subsection{Ball Bearings}

\begin{figure}[ht]
\begin{center}
   \includegraphics[width=0.8\linewidth]{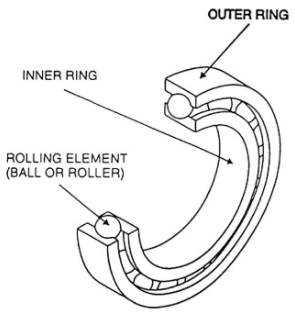}
\end{center}
   \caption{This diagram shows the 3 main components of a bearing joint: the inner race, the outer race, and the ball bearing or rolling element [2].}
   \label{bearing joint diagram}
\end{figure}

Bearing joints are the principal component of rotating machinery. Figure \ref{bearing joint diagram} shows a schematic of the main components of the joint. When the joint works properly, the ball bearings or rolling elements are free to rotate between the inner and outer races. This allows the inner and outer races to spin independent of one another, creating a rotating joint.

A fault is introduced to the system when there is a ridge, bump, or some other abnormality anywhere along the surface of one of the three main components: the ball, the inner race, or the outer race. These faults can be of varying sizes and are often the result of a manufacturing defect or overuse of the joint. The focus of the experiments is to identify whether there is a fault in the joint and locate which of the three main components it is on.

The primary focus of this project is to recreate and validate the results used in ``Rolling Element Bearing Fault Diagnosis Using Convolutional Neural Network and Vibration" by Hoang and Kang, which preprocesses the signals using a new technique and trains convolutional neural network to detect the presence of a fault and locate which of the three main components it is on [3]. The Related Work section explains the data used in the research and describes in detail the construction of vibration images. The use of vibration images is the primary focus of the results being recreated, and the limitations of this method are discussed in this section as well. The Methods section defines the convolutional neural network architecture used in the experiments and discusses data augmentation techniques used to balance the number of samples for each class or fault location in the data. The experiments sections contains data collected from 2 experiments. The first experiment is modeled from Hoang and Kang's paper being recreated. It applies the related methods to a data set 20 times larger than done previously in order to demonstrate the robustness of vibration images to bearing faults of different sizes in motors run at different horsepower. The second experiment is designed to highlight the hyperparameters used in vibration image construction, a point which was lacking in Hoang and Kang's paper. This is done by performing a similar procedure but using vibration images constructed from sampling vibration signals at different frequencies. The goal of both of these experiments is to validate vibration images as a valid preprocessing technique and highlight potential areas for future research.


\section{Related Work}

\subsection{Case Western Reserve Dataset}

The Case Western Reserve Ball Bearing Data Center (CWR) is a publicly available, benchmark dataset for ball bearing IFD [4]. It was generated using an electric motor which contained a bearing joint. Faults of different sizes (7, 14, and 21 mils) were applied one at a time to all three parts of the bearing joint. As the motor was run, vibration signals were recorded. This experimental setup was conducted with multiple fault sizes for each location and at multiple horsepowers (hp).

\subsection{Vibration Images}

Vibration images are a technique for analyzing the CWR 1D vibration signals [3]. Hoang and Kang define a vibration image with equation \ref{vibration image equation}:

\begin{equation}
    \label{vibration image equation}
    VI[i,j] = S[(i-1)\times l + j]
\end{equation}

\begin{figure*}
\begin{center}
\includegraphics[]{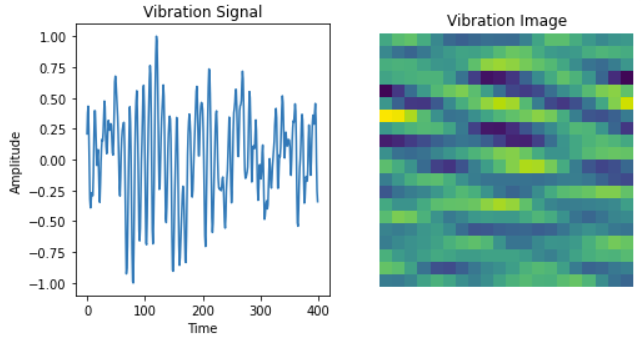}
\end{center}
   \caption{The vibration signal on the left has been normalized to $[-1,1]$ and its corresponding vibration image is seen on the right. The signal has 400 samples in it, each of which corresponds to a specific pixel in to $20\times 20$ vibration image on the right based on equation \ref{vibration image equation}.}
\label{vibration signal and image}
\end{figure*}

In equation \ref{vibration image equation}, $VI$ is the vibration image, and $S$ is the signal. This mapping from 1D to 2D data can be described as taking the signal and splitting it into $l$ segments of $l$ samples each, where each segment represents the intensity values of one row of the vibration image. In principle, the dimensions of the image are not required to be equal, but the initial formulation constructed all vibration images as squares. Keeping with this, the mapping $S\xrightarrow{}VI$ is parameterized by $l$. In the paper $l=20$ was chosen [3]. Figure \ref{vibration signal and image} shows a signal and its corresponding vibration images.

Using vibration images, convolutional neural networks (CNNs) were trained to detect and locate faults. These CNNs achieved 100\% accuracy in terms of detecting and locating faults in bearing joints using the CWR data with the fault size set to 7 mils and the motor run at 0 hp [3].

\subsection{Limitations of Vibration Image}

While Hoang and Kang present an interesting method of preprocessing data and demonstrate a very high performance of the model, there are two areas where their work could be improved upon. First off, reporting 100\% accuracy using a deep learning classifier is a cause for concern based on one of two potential reasons: either ball bearing IFD is not sufficiently complicated and can easily be solved without a CNN, or the model is being trained and tested on a data whose scope is too narrow. Given that IFD is an active field of research and the 7 mil, 0 hp data only accounts for approximately 5\% of the CWR dataset, it seemed likely that the model was overfit to the data. Extending on their work, a main result presented in our experiments section applies the aforementioned vibration image construction technique and an identical CNN architecture to the entire CWR dataset to validate the robustness of Hoang and Kang’s methods.

For the second improvement, the mapping of the signal to the vibration image defined in equation \ref{vibration image equation} was left relatively unjustified and appeared inconsistent with the use of a CNN. Using a vibration image of size $20\times 20,$ which is constructed from 400 samples of the vibration signal, the first kernel of size (5,5) in the CNN will take samples 1-5, 21-25, 41-45, and 51-55 from the signal as inputs. The relationship between these sampling points and reason they should be considered in the same kernel is not clear.

The idea for CNNs was initially introduced for the purpose of position invariant pattern recognition [5]. Rather than position data, the vibration signals are recorded in time. It makes sense to think of sample points above as points in time. This requires knowledge of the sampling rate of the vibration signal, otherwise signals sampled at different frequencies would not be comparable. A $20\times 20$ vibration image made from a signal sampled at 400 Hz represents 1 second but from a signal sampled at 100 Hz represents 4 seconds, so these images shouldn't necessarily be comparable. It follows from this that the optimal parameterization of the size of each vibration image may be a function of the sampling frequency of the image as well as a parameter $l$. This point is not explored by Hoang and Kang, but seems fundamental to vibration image construction.


\section{Methods}

\begin{figure*}[]
\begin{center}
   \includegraphics[width = 0.8\linewidth]{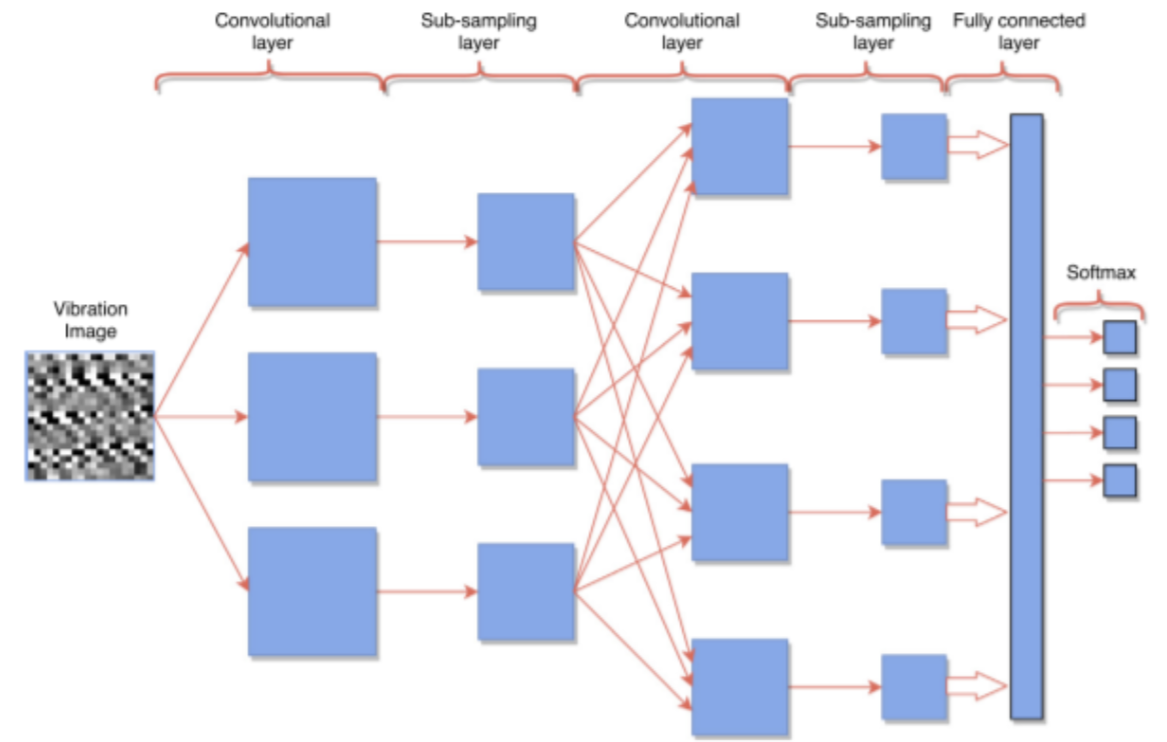}
\end{center}
   \caption{This is the architecture of the neural network used in all of our experiments. This figure and architecture is identical to the one used in the previous work with vibration images [3].}
   \label{CNN architecture}
\end{figure*}

The focus of our experiments was to apply the process described above to investigate robustness of the model to a larger dataset and different sampling frequencies. Using the drive end of the motor vibration signals from CWR, the signals are normalized to the range [-1,1] and vibration images were constructed. The CNN architecture parallels that from the paper, which consisted of a 2D convolution layer with kernel size (5,5),  a subsampling layer, a second 2D convolutional layer with kernel size (3,3), a second subsampling layer, a fully connected layer with a ReLu activation function, and a Softmax output layer. The CNNs were trained with 150 epochs, batch sizes of 50, and used stochastic gradient descent with a learning rate of 0.01 and momentum of 0.9. This architecture is shown in figure \ref{CNN architecture}.

In the cases where there was significant data imbalance which occurred when multiple fault sizes were used, we employed data augmentation methods to rectify the imbalances. The vibration images representing baseline data (no faults) were rotated to balance out the class distribution. Only the baseline cases were augmented because CWR has the least data for healthy bearing joints.

\section{Experiments}

All experiments consist of applying the vibration image construction and training the CNN architecture using the process as described in the methods section. To ensure the robustness of the method each experiment was run five times so that the standard deviation of the performance could be calculated, and during each run a random $\frac{5}{6}$ of the data was reserved for training the model and the remainder was used for testing the model's accuracy.

\subsection{Vibration Image Robustness}

The Haong and Kang model recorded 100\% accuracy using only data with 7 mil faults and the machine set to 0 hp. After running their experiments on larger sets of data, table \ref{experiment 1 results} shows the accuracies recorded from multiple partitions of CWR.

\begin{table}[ht]
    \centering
    \begin{tabular}{|c|c|c|}
        \hline
        Data&Accuracy&Standard Deviation\\
        \hline
        7 mil faults, & 92.8 & 0.9\\
        0 hp (n=2,400)& & \\
        \hline
        All faults, 0& 75.1 & 1.2\\
        hp (n=6,800) & & \\
        \hline
        7 mil faults, 0-3& 97.4 & 1.9\\
        hp (n=17,000)& & \\
        \hline
        All faults 0-3 hp& 86.6 & 0.4\\
        (n=50,400) &&\\
        \hline
    \end{tabular}
    \caption{The results in this table demonstrate that using vibration images as a preprocessing technique works effectively on a wide range of ball bearing data.}
    \label{experiment 1 results}
\end{table}

Comparing replicated results from Hoang and Kang, we had 92.8\% compared to their 100\%, a difference we did not consider to be significant. Given that their training parameters were not published, it is reasonable to believe that 100\% accuracy could be achieved. From the table above, it seems that our model generalizes well at different hp, and the increase in accuracy may be explained by increase in training data. However, the model performs less well under different fault sizes, which is problematic for IFD when fault specifications are unknown, which is commonly the case in industry. When all of the data is used in the fourth row, the accuracy increases again, likely from the increased training data.

\subsection{Frequency Analysis of Vibration Images}

To address the question of optimal vibration image construction, the vibration signals are subsampled to produce signals sampled at different frequencies. The initial signals are sampled at a rate of 48k Hz [4]. Using these subsampled signals, vibration images of size $20\times 20$ are constructed and the model is trained as described previously. Table \ref{experiment 2 results} shows the results of training the model on these vibration images.

\begin{table}[ht]
    \centering
    \begin{tabular}{|c|c|c|}
        \hline
        Sampling Frequency&Accuracy&Standard Deviation\\
        \hline
        48,000 Hz&92.8&0.9\\
        \hline
        24,000 Hz&94.0&2.2\\
        \hline
        16,000 Hz&77.5&3.1\\
        \hline
        12,000 Hz&86.2&5.5\\
        \hline
    \end{tabular}
    \caption{This table shows the performance of a CNN trained on vibration images sampled at different frequencies. There is not a linear relationship between the sampling rate and the models performance.}
    \label{experiment 2 results}
\end{table}

The increase in standard deviation of the model’s performance as the sampling frequency changes is likely the result of decreased training data. However, the significant decrease in accuracy is not a linear function of the sampling frequency, which is apparent because the model has worse performance when the data is sampled at 16k Hz. While this doesn’t capture the entire dynamic between the size of the images and the sampling frequency, it does suggest that sampling frequency is another hyperparameter to consider when creating vibration images. This point was not made by Hoang and Kang and may be the focus of future research.

\section{Conclusion}

Constructing vibration images is a new technique for signals processing that has the potential to capture features in a signal that may otherwise not be apparent. Through this paper, results from Hoang and Kang were validated and their methods were applied to the remainder of CWR to achieve satisfactory detection results, demonstrating that this technique is reasonably robust.

While the work in the second experiment doesn't prescribe the optimal sampling frequencies or image size for a signal, it does indicate that these are significant hyperparameters that must be considered for vibration image construction. Vibration images have the potential to be highly effective in signals that have defined morphologies which could be aligned in each row of the image, such as aligning the QRS complex of an ECG in each row of the image. This would solve the problem of vibration image size and sampling frequency, and it is an area for potential future research. The logical next step for researchers could be to build upon fine-tuning vibration image construction, and apply IFD to more applications beyond machinery. To conclude, we have shown that our version of IFD (and ones similar to it) can be reasonably robust, sampling from a general, less biased dataset, performing well using techniques from previous research. 

\textbf{Note:} This work deviated from our original proposal in that it focused more on investigating vibration images, expanding the data used, and validating the use of them than it did comparing them to other standard 1D signals processing techniques and filtering schemes, as we initially proposed. 

\section*{References}

\begin{flushleft}
[1] S. Nandi, H. A. Toliyat and X. Li, ``Condition Monitoring and Fault Diagnosis of Electrical Motors—A Review," in IEEE Transactions on Energy Conversion, vol. 20, no. 4, pp. 719-729, Dec. 2005, doi: 10.1109/TEC. 2005.847955
D.-T. Hoang and H.-J. Kang, “Rolling element bearing fault diagnosis using convolutional neural network and Vibration Image,” Cognitive Systems Research, vol. 53, pp. 42–50, Mar. 2018. 
\end{flushleft}
\begin{flushleft}
[2] M. Forsthoffer, “Journal (radial) bearings,” \textit{Forsthoffer's Component Condition Monitoring}, pp. 45–51, 2019. 
\end{flushleft}
\begin{flushleft}
[3] D.-T. Hoang and H.-J. Kang, “Rolling Element Bearing Fault Diagnosis Using Convolutional Neural Network and Vibration Image,” Cognitive Systems Research, vol. 53, pp. 42–50, Mar. 2018. 
\end{flushleft}
\begin{flushleft}
[4] “Bearing Data Center: Case School of Engineering: Case Western Reserve University,” Bearing Data Center, 11-Aug-2021. [Online]. Available: https://engineering. case.edu/bearingdatacenter. [Accessed: 30-Nov-2021].
\end{flushleft}
\begin{flushleft}
[5] K. Fukushima, “Neocognitron: A Self-organizing Neural Network Model for a Mechanism of Pattern Recognition Unaffected by Shift in Position,” 08-Oct-1979. [Online]. Available: https://www.rctn.org/bruno/public/ papers/Fukushima1980.pdf. [Accessed: 06-Dec-2021].
\end{flushleft}

\end{document}